\documentclass{ecai} 

\usepackage{latexsym}
\usepackage{amssymb}
\usepackage{amsmath}
\usepackage{amsthm}
\usepackage{booktabs}
\usepackage{enumitem}
\usepackage{graphicx}
\usepackage{color}
\usepackage{hyperref}
\usepackage{cleveref}
\usepackage{subcaption}
\usepackage[aboveskip=6pt, belowskip=13pt]{caption}

\newcommand{\BibTeX}{B\kern-.05em{\sc i\kern-.025em b}\kern-.08em\TeX}

\begin{document}


\begin{frontmatter}




\title{From Competition to Centralization: \\ The Oligopoly in Ethereum Block Building Auctions}


\author[A]{\fnms{Fei}~\snm{Wu}\thanks{Corresponding Author. Email: fei.wu@kcl.ac.uk}}
\author[B]{\fnms{Thomas}~\snm{Thiery}}
\author[A]{\fnms{Stefanos}~\snm{Leonardos}}
\author[A]{\fnms{Carmine}~\snm{Ventre}}

\address[A]{Department of Informatics, King's College London}
\address[B]{Ethereum Foundation}


\begin{abstract}

Block production on the Ethereum blockchain has adopted an auction-based mechanism known as \emph{Proposer-Builder Separation} (PBS), where validators outsource block creation to \emph{builders} competing in \emph{MEV-Boost auctions} for Maximal Extractable Value (MEV) rewards. We employ empirical game-theoretic analysis based on simulations to examine how advantages in latency and MEV access shape builder strategic bidding and auction outcomes. We find that a small set of dominant builders leverage these advantages, consolidating power, reducing auction efficiency, and heightening centralization. Our results underscore the need for fair MEV distribution and sustained efforts to promote decentralization in Ethereum’s block building market.
\end{abstract}

\end{frontmatter}


\section{Introduction}
Modeling and analyzing strategic behavior in multi-agent systems is crucial for understanding how incentives drive agents’ decisions and shape overall system outcomes. Many applications rely on equilibrium analysis to predict how self-interested agents will act under various constraints. However, the realism of these predictions hinges on the breadth of agents’ strategy spaces: when strategies are constrained to be narrow, hand-crafted heuristics, interactions can be oversimplified, resulting in findings that fail to generalize or match real-world observations.

In this paper, we adopt the \emph{meta game} framework \cite{metagame} to capture a broader set of agent behaviors, defined by \emph{meta strategies} \cite{metastrategy}, and thus, model multi-agent decision-making more comprehensively. This approach allows agents to explore a wider range of strategies, uncovering behaviors that might remain hidden when relying solely on pre-defined or empirically derived heuristics. We demonstrate the effectiveness of this framework in the context of Decentralized Finance (DeFi), specifically the mechanism \emph{Proposer-Builder Separation} (PBS) and the Ethereum block building market \cite{buterin2014eth,pbs}—while emphasizing that our methodology is broadly applicable to other multi-agent environments.

DeFi provides a rich case study. In Ethereum’s current block production process, PBS decouples the block proposal and block building processes, allowing the proposer—a validator selected by the protocol to add a new block—to outsource the task of block building and MEV extraction to specialized entities known as \emph{builders} \cite{pbs}. Profit-driven builders compete in English-style auctions, known as \emph{MEV-Boost auctions} \cite{mevboost}, for the right to produce new blocks and earn potential \emph{Maximal Extractable Value} (MEV) rewards \cite{mev}. Ideally, each builder would have equal network access and resources, creating a fair, competitive market with an efficient auction. In practice, however, builders seek to secure advantages—such as exclusive private transactions (orderflow) and reduced latency—to gain a competitive edge. Consequently, as of April 2025, two builders (\texttt{beaverbuild} and \texttt{Titan}) dominated over 90\% of the auctions \cite{onchaindata,mevboostpics}, reaping high profits and raising centralization concerns.

Despite empirical studies revealed such oligopoly in the Ethereum block building market \cite{yang2024decentralization, whowinsandwhy}, insights into how latency and private orderflow asymmetries shape bidding incentives in MEV-Boost auctions remain limited. Prior work restricts builders to empirically validated or narrowly adaptive heuristic strategies \cite{wu2024strategic,icaifpaper}, yielding limited insight into emerging behaviors and incentives. Counterintuitively, \cite{icaifpaper} finds that builders exhibit mild collusion that leads to price-fixing, even with private orderflow asymmetries. 

In this paper, we address this gap by considering MEV-Boost auctions within the meta-game framework that broadens the strategy space, capturing more comprehensive bidding behaviors (Section~\ref{sec:model}). We investigate builders' strategic bidding incentives under various scenarios (Section~\ref{sec:games}) and show that advantages in latency and private orderflow can significantly shift bidding incentives and auction outcomes. These results align more closely with real-world observations and offer novel insights into the key dynamics and builders' incentives that drive the outcomes in current MEV-Boost auctions (Section~\ref{sec:results}). In particular, our paper makes the following contributions. \par
First, we find that, under idealized conditions, where builders have similar latency and orderflow access, builders are incentivized to compete against each other by bidding a significant portion of block value, resulting in an equal market share (win rate). This fosters a decentralized and competitive Ethereum block building market. \par
Second, we identify which asymmetry contributes to centralization. Compared to the previous idealized scenario, our results indicate that the latency advantage alone within certain threshold does not significantly alter builders' bidding incentives or contribute to centralization. However, latency asymmetry can lead to reduced proposer revenue (i.e., lower winning bid values) and decreased auction efficiency. These two findings contrast sharply with those in \cite{icaifpaper}, highlighting the importance of strategy space design.\par
Third, we demonstrate that advantage in private orderflow access significantly impacts builders' bidding incentives. Builders with greater access to private orderflow are incentivized to bid a smaller portion of the block value while dominating the market and maintaining higher profit margins. Conversely, builders with limited orderflow access are forced to bid aggressively. This disparity in orderflow access leads to centralization, oligopolistic behaviors, and auction inefficiencies in the Ethereum block building market.\par
Finally, we conclude our analysis by discussing the broader challenges in the Ethereum block building market and the implications of our approach and analysis for future studies (Section~\ref{sec:discussion}). 

\section{Background and Related Works}
\paragraph{Ethereum blockchain:} The Ethereum blockchain grows by one block every 12-second \emph{slot}, as specified by its Proof-of-Stake consensus protocol \cite{pos}. At each slot, the protocol randomly selects a validator to act as the proposer, responsible for adding a new block to the blockchain at the start of that slot \cite{honestvali}.

\paragraph{MEV supply chain and vertical integration:} MEV refers to the total value extractable from a blockchain state through transaction inclusion, exclusion, and ordering \cite{flashboys}, which are often carried out by MEV searchers under the current PBS framework with varying strategies \cite{qin2022quantifying,defiliquidations,obadia2021unity,wu2025measuringcexdexextractedvalue}. The MEV supply chain outlines this process: Searchers share extracted MEV with builders by offering tips or payments for prioritized transaction inclusion, while builders share part of their MEV profits with proposers through bidding in the MEV-Boost auction. Vertical integration occurs when builders optimize this process by directly controlling searcher activity and relay infrastructure to gain an edge in MEV-Boost auctions with advantages in latency and access to searcher orderflow.

\paragraph{PBS and MEV-Boost:} Although Proposer-Builder Separation (PBS) has not yet been formally integrated into the Ethereum consensus protocol, \emph{MEV-Boost} \cite{mevboost} was developed as an out-of-protocol solution to enhance decentralization. MEV-Boost introduces \emph{relays}, trusted third-party intermediaries, to facilitate a fair exchange between the proposer and builders. Relays ensure that the proposer receives the payment from the winning builder, while builders are protected from having their MEV opportunities revealed and stolen. The proposer can either produce the block locally and extract MEV themselves, or use MEV-Boost to outsource block production and earn MEV revenue from builders. MEV-Boost auctions align with the 12-second slot interval: to get their block selected by the proposer of slot $n$, builders begin to submit their blocks along with bids to relays at the start of slot $n-1$. Relays validate these bids and make them available to the proposer, who typically terminates the auction by signing the block header with the highest bid and returning it the to corresponding relay, which then publishes the full block to the network.

\paragraph{Related Works.} Apart from \cite{icaifpaper}, previous research on builders' behaviors in MEV-Boost auctions has explored various empirical, theoretical, and game-theoretic perspectives. \cite{wahrstatter2023time, heimbach2023ethereum} conducted a longitudinal empirical study, revealing a positive correlation between bid timing and value in block production and highlighting centralization concerns in the Ethereum builder market. \cite{cancellation} identified strategic behaviors like bid erosion and bid shielding using empirical data, while \cite{bbp} confirmed diverse bidding patterns among builders. \cite{whowinsandwhy} empirically highlighted key factors behind successful block wins, such as exclusive access to private orderflow providers. \cite{yang2024decentralization} empirically confirmed that private orderflow serves as a significant market entry barrier for new builders, impacting efficiency and competitiveness in the market. \cite{gupta2023centralizing} theoretically demonstrated that builders with high-value orderflow tend to perform better in MEV-Boost auctions and highlighted the centralizing effect of private orderflow in the Ethereum block building market. These findings were also empirically validated by \cite{cefidefiofa}. \cite{pai2023structural} further proved that latency advantages give builders a competitive edge. \cite{wu2024strategic} introduced a game-theoretic model alongside exemplary bidding strategies showcasing the importance of both latency and orderflow access with simulations.

\section{Model}
\label{sec:model}

The MEV-Boost auction was initially modelled in \cite{wu2024strategic} and \cite{icaifpaper}. In the current study, we enrich this model with the introduction of \emph{meta strategies} and the construction of a \emph{meta game} that better captures the practical aspects of the auction and the builders' strategic decisions. To keep the paper self-contained, we first recap the main model's components and parameters. \par

\paragraph{MEV-Boost auction:} We consider a set of $N = \{1, \ldots, n\}$ \emph{builders} or \emph{players} competing in the MEV-Boost auction game. Each builder, indexed by $i$, employs a bidding strategy $s_i: X \to \mathbb{R}_+$, where $x_{i,t} \in X$ represents a vector of \emph{input variables} at time $t \geq 0$. When contextually clear, we will omit the dependence of $s_i$ and $x_{i,t}$ and write $s_{i,t}$ to denote the bid of player $i$ at time $t$.\par
The input variables typically consist of the public and private signals. The \emph{public signal}, $P(t)$, represents the total extractable value (MEV) from public transactions in the mempool at time $t$. The private signal, $E_i(t)$, captures the MEV from private (or exclusive) orderflow secured by builder $i$. Since builders may share portions of private orderflow, we introduce an orderflow access probability, $\theta_i \in [0,1]$, representing the likelihood that builder $i$ accesses a given orderflow. These probabilities remain constant throughout the auction. Both the public signal, $P(t)$, and private signal, $E_i(t)$, are driven by stochastic processes. The number of public transactions, $N(t)$, follows a Poisson process with rate $\lambda_p$, and each transaction value, $V_j$, is drawn from a log-normal distribution. Similarly, the number of private orderflows, $N_i(t)$, accessed by builder $i$, follows a Poisson process with rate $\lambda_e \cdot \theta_i$, with each orderflow value, $O_j$, also drawn from a log-normal distribution. Thus, the public and private signal of player $i$ at time $t$ are
\begin{align}
P(t) := \sum\nolimits_{j=1}^ {N(t)} V_j, \text{ and }
E_i(t) := \sum\nolimits_{j=1}^{N_i(t)} O_j.
\end{align}
We further let $E(t)$ denote the sum of all private orderflow values at time $t$. By combining the public signal and the private signals, we obtain the \emph{aggregated signal} for player $i$, $L_i(t):= P(t) + E_i(t)$, and the \emph{total signal} in the auction (i.e., the total MEV available), $L(t):= P(t) + E(t)$, at time $t$.

\paragraph{Latency and current highest bid:} The \emph{latency}, $\Delta_i>0$ of builder $i$ quantifies the delay in the relay's acceptance of bids relative to that builder's access to a signal update and their subsequent bid submission. It mostly depends on that builder's connectivity to orderflow providers and relays. It is assumed to be known and constant during the auction and to only affect the player's bidding action. The \emph{current highest bid}, denoted by $\max_{j \in N}\{s_{j,k}\}_{k \leq t}$, represents the 
highest bid 
accepted by the relay up to time $t$.

Finally, the \emph{auction interval} is defined as $[0, T]$, where $T$ denotes the time when the proposer selects the highest winning bid. Instead of $T$ being exactly equal to 12 seconds as expected, the winning bid is typically selected around $T = 13$ seconds according to empirical data \cite{mevboostpics, latencyismoney} due to factors such as the proposer's latency or strategic behaviors \cite{schwarz2023time}. Thus, we draw $T$ randomly from a normal distribution with mean $13$ and standard deviation $\sigma$.\footnote{We set $\sigma=0.1$ in simulations but the results are analogous across different values.}

\subsection{Model Calibration}
We implement the model and simulate the players' bidding behaviors using \emph{Agent-Based Modeling} (ABM) techniques. Due to technical constraints, time progresses in discrete steps of 10ms. At each time step, players submit their bids simultaneously, and the auction state is then updated accordingly.

We calibrate the model against 75 days of mempool data \cite{mempooldata} and on-chain data \cite{onchaindata} maintained by Flashbots and Dune Analytics covering the period from July 11 to September 23, 2024. Specifically, we consider $n=10$ players in the MEV-Boost auction game, as the top 10 builders build 97.51\% of all blocks built via MEV-Boost. Each block contains approximately 105 public transactions, with the MEV of each transaction valued at approximately 0.00019 ETH, collectively representing nearly 31\% of the total block value.

\paragraph{Estimating private orderflow:} It is important to recognize a degree of bias in data related to private orderflow. On-chain data only reveals the private orderflow included by the winning builder. Since this builder wins the auction, we infer that their private orderflow access exceeds that of their competitors. However, a builder's actual private orderflow access remains unknown, as this data is neither recorded on-chain nor available through any Relay Data API.\par
As a result, we assume that private orderflow access among builders is randomly distributed within a domain that we estimate from on-chain data \cite{onchaindata}. This ensures that the volume of private orderflow in the winning block matches expectations from on-chain metrics. For the top 10 builders, private orderflow accounts for 10.92\% (by builder \texttt{blockbeelder}) to 36.38\% (by builder \texttt{Titan}) of transactions in their winning blocks. Accordingly, we infer that 36.38\% represents the maximum possible private orderflow, corresponding to 100\% access probability, while 10.92\% marks the minimum, equating to a 30\% probability. Thus, we model private orderflow access probabilities, $\theta_i$, as uniformly distributed between 30\% and 100\% (equivalently $[0.3,1]$).

\subsection{Meta-game and meta-strategies}
To prevent proposers from locally building blocks using public mempool transactions, builders are incentivized to bid at least the value of the public signal \cite{worthusingmevboost}. Consequently, the builder's strategic decision is how much of their private signal to include in their bid at time $t$. However, since players can bid any portion of their private signal, the strategy profile becomes infinite, making this game intractable.\par
To address this complexity, we introduce a \emph{meta game} \cite{metagame}, which simplifies the original game by focusing only on a set of \emph{meta strategies} that summarise all possible strategies \cite{metastrategy}. In the context of MEV-Boost auctions, we define the meta strategy set as $S = \{\emph{"conservative", "moderate", "aggressive"}\}$, or $S = \{m_1, m_2, m_3\}$. In particular, we divide the $[0,1]$ interval, representing the fraction of private signal included in a bid, into three sub-intervals that correspond to these strategies. Thus, the bid of a builder, $i$, is given by (cf. \Cref{tab:bidding_strategies})
\begin{equation}\label{eq:meta}
    s_i(x_{i,t},m):= P(t)+\lambda_i(m) \times E_i(t)
\end{equation}
where, for each auction simulation, the exact fraction, $\lambda_i(m)$, of private signal is drawn uniformly from the corresponding sub-interval based on each builder's meta-strategy choice $m\in S$, i.e., $[0,1/3]$ for $m_1$, $(1/3, 2/3]$ for $m_2$, and $(2/3, 1]$ for $m_3$.\par
The above definitions assume that builders are not willing to bid in a way that leads to a negative profit. However, in current practice, builders sometimes subsidize their bids to win the auction to maintain a market share and expect better performance in the future \cite{yang2024decentralization, whowinsandwhy}. We introduce this assumption here to simplify the analysis, focusing on non-repeated auction games.\par

Each builder $i\in N=\{1,\dots,10\}$ selects a pure meta-strategy $m_i \in S$, and bids according to that throughout the auction interval. The payoff, $u_i(\mathbf{m})$, of builder $i$ at the meta-strategy profile, $\mathbf{m} = (m_1, m_2, \dots, m_{10})$, is given by
\begin{equation}\label{eq:utility}
u_i(\mathbf{m}) :=
\begin{cases}
L_i(t_w) - s_{i,t_w} & \text{if }s_{i,t_w}=\max_{j \in N}\{s_{j,k}\}_{k \leq T}, \\
0 & \text{otherwise}.
\end{cases}
\end{equation}
where $t_w$ is the submission time of the winning bid, and as usually, $s_{i,t_w}=s_i(x_{i,t_w},m_i)$ is the bid of builder $i\in N$ at $t_w$.


\begin{table}[t]
\centering
\caption{Meta strategies and bid values $s_i(x_{i,t},m_i).$}
\vspace{-0.1cm}
\par
\label{tab:bidding_strategies}
\begin{tabular}{@{}l@{\hspace{3pt}}l@{\hspace{9pt}}l@{\hspace{9pt}}l@{}}
\hline
\multicolumn{2}{@{}l}{\textbf{meta-strategy}} & \textbf{
bid value at $t\leq T$} & \textbf{private-signal fraction} \\\hline
$m_1$ & conservative & $P(t) + \lambda(m_1)\cdot E_i(t)$ & $\lambda(m_1)\sim U[0,1/3]$ \\ \hline
$m_2$ & moderate & $P(t) + \lambda(m_2)\cdot E_i(t)$ & $\lambda(m_2)\sim U(1/3,2/3]$\\ \hline
$m_3$ & aggressive  & $P(t) + \lambda(m_3)\cdot E_i(t)$ & $\lambda(m_3)\sim U(2/3,1]$\\ \hline
\end{tabular}
\end{table}

\section{Game Settings}
\label{sec:games}
While introducing a meta game effectively reduces the size of the original game, explicitly solving the meta game is still computationally challenging due to the large strategy space: with 10 asymmetric players each choosing from 3 meta strategies, the number of distinct strategy profiles reaches $3^{10}$. To mitigate this complexity, we further reduce the size of the game by leveraging certain symmetries among players and adopting an empirical game-theoretic approach. Specifically, players are characterized by two key attributes: their latency and their private orderflow access probability. Based on these attributes, we analyze three different game variants, each assuming different levels of uniformity between players.

This section introduces the game definitions, discusses payoff representations, and explains methods used to solve these games.

\subsubsection*{Uniform latency and orderflow access distribution} We first consider a scenario where all players exhibit the same latency and have their private orderflow access probabilities drawn from the same prior distribution. Specifically, in each auction simulation, for each player $i$, $\theta_i$ is sampled uniformly on $[0.3, 1]$.

In this way, we can simplify the analysis by transforming the game into an \emph{anonymous game}, where a player's payoff depends only on the number of opponents adopting each strategy \cite{wellman2024survey}. This approach allows us to represent a strategy profile as a vector containing the count of players choosing each strategy, thereby reducing the number of profiles from $3^{10}$ to $\binom{10 + \lvert S \rvert - 1}{10} = 66$.

We then replace the original payoff matrix with a \emph{heuristic payoff table} (HPT) \cite{2002hpt, metagame}, where the payoffs of each meta strategy are stored as a function of the number of players using it. However, despite the players' probabilities being drawn from the same prior distribution, their realized access may differ, leading to variability in payoffs even for identical strategy choices. This inherent asymmetry complicates the application of standard HPT methods, which assume symmetry across players \cite{metagame}.

To address this, we define the payoff for each strategy as the \emph{average payoff} earned by all players adopting that strategy within a given profile. Additionally, we reduce the impact of random fluctuations in orderflow access as well as the portion of private signal associated with each meta strategy by averaging the payoffs across 1,000 auction simulations for each strategy profile. This approach effectively makes the game \emph{symmetric} by leveraging the fact that, in expectation, players with the same latency and strategy tend to achieve the same payoff due to their access probabilities being drawn from the same distribution.

Formally, we define the HPT as $\mathcal{H} = (\mathcal{N}, \mathcal{U})$, where $\mathcal{N}$ is a $\binom{10 + \lvert S \rvert - 1}{10} \times \lvert S \rvert$ matrix that encodes the strategy profiles, and $\mathcal{U}$ is a matrix of the same dimension that represents the corresponding payoff. Each row $\mathcal{N}_k$ in $\mathcal{N}$ corresponds to a meta strategy profile $m^k$. The element $\mathcal{N}_{k,j}$ in $\mathcal{N}$ describes the count of players choosing strategy $m_j \in S, j \in \{1,2,3\}$ in strategy profile $m^k$. Correspondingly, the element $\mathcal{U}_{k,j}$ in $\mathcal{U}$ describes the average payoff of players choosing strategy $m_j$ in profile $m^k$. $\mathcal{U}_{k,j}$ can be given by
\begin{equation}
\mathcal{U}_{k,j} =
\begin{cases}
\frac{1}{\mathcal{N}_{k,j}} \sum_{i: m_i = m_j} u_i\left(m^k\right) & \text{if } \mathcal{N}_{k,j} > 0, \\
0 & \text{otherwise}.
\end{cases}    
\end{equation}

\begin{table}[t]
\centering
\caption{Game settings.}
\vspace{-0.1cm}
\label{tab:empirical_games}
\begin{tabular}{@{}l@{\hspace{20pt}}l@{\hspace{20pt}}l@{}}
\hline
\textbf{game} & \textbf{latency} & 
\begin{tabular}[t]{@{}l@{}}\textbf{orderflow access}\\\textbf{probability}\end{tabular}
\\\hline
symmetric & same & same distribution\\
role-symmetric & \begin{tabular}[t]{@{}l@{}}low ($5$) or high ($5$)\\\end{tabular} & same distribution\\
role-symmetric & same & low ($5$) or high ($5$)
\\ \hline
\end{tabular}
\end{table}

\subsubsection*{Heterogeneous latencies} In the previous game setting, we assumed that all players had uniform latency. However, in practice, builders often experience different latencies \cite{bbp, whowinsandwhy}. In this section, we consider a scenario where players share the same distribution for private orderflow access but differ in their latencies. Specifically, we model two distinct latency groups: one group consisting of 5 players with low latency and another group consisting of 5 players with high latency.

This more realistic scenario allows us to model the game as a \emph{role-symmetric game} \cite{wellman2024survey}, where players are categorized into \emph{roles} based on their latency, and their payoffs depend on the strategies used within their role. In this case, there are two roles: $r_l$ for the low-latency group and $r_h$ for the high-latency group. 

We denote the role of each player $i$ as $r_i \in \{r_l, r_h\}$. We then extend the previous HPT structure to accommodate this role symmetry. Specifically, let $\mathcal{M} = (\mathcal{N}^l \times \mathcal{N}^h , \mathcal{U}^l \times \mathcal{U}^h)$. $\mathcal{N}^l \times \mathcal{N}^h$ is a $\binom{5 + \lvert S \rvert - 1}{5}^2 \times 2\lvert S \rvert$ matrix of strategy profile representations incorporating the roles, where $\mathcal{N}^l$ and $\mathcal{N}^h$ are matrices of the count of players adopting each strategy in the low-latency and high-latency groups, respectively. The entry $\mathcal{N}^r_{k,j}$ in $\mathcal{N}^r$, for $r \in \{r_l, r_h\}$, describes the number of players choosing strategy $m_j \in S, j \in \{1,2,3\}$ within role $r$ in the strategy profile $m^k$, and the corresponding entry $\mathcal{U}^r_{k,j}$ in $\mathcal{U}^r$ describes the average payoff of players within the role $r$ choosing the strategy $m_j$ in the profile $m^k$. $\mathcal{U}^r_{k,j}$ can be given by

\begin{equation}
    \mathcal{U}^r_{k,j} =
\begin{cases}
\frac{1}{\mathcal{N}^r_{k,j}} \sum_{i: m_i = m_j, r_i = r} u_i\left(m^k\right) & \text{if } \mathcal{N}^r_{k,j} > 0, \\
0 & \text{otherwise}.
\end{cases}
\end{equation}

\subsubsection*{Different private orderflow access probabilities.} The final variant considers differences in orderflow access probabilities among builders. In practice, private orderflow access is not uniform across all players; some builders have higher probabilities of accessing valuable orderflow than others \cite{whowinsandwhy}. In this variant, we model the game with 5 players having high private orderflow access probabilities and 5 players having low access probabilities, while all players have the same latency.

Similar to the latency-based setting, this scenario is also modeled as a role-symmetric game, with players divided into two roles based on their private orderflow access: one role for players with high access and one for those with low access. The payoff structure is handled similarly, accounting for the differences in private orderflow access among players.

\subsection{\texorpdfstring{$\alpha$-Rank}{Alpha-Rank}}

To solve the games discussed above, we utilize the \emph{$\alpha$-Rank} algorithm \cite{alpharank}. The $\alpha$-Rank algorithm models a stochastic evolutionary process to capture the selection-mutation dynamics of a population's strategy choices. This method provides a dynamic solution to the game by understanding the agents' behaviors and predicting their convergence towards equilibrium. The solution (equilibrium), presented as a ranking of strategy profiles, corresponds to the stationary probabilities, also referred to as the \emph{mass}, within the unique stationary distribution of the $\alpha$-Rank Markov Chain, which captures the evolutionary stable probabilities of the strategy profiles. We denote the stationary distribution as $\pi$ and the mass of a strategy profile $m^k$ as $\pi_k$. To convey the equilibrium clearly, we represent it as the average number of players using each strategy across all profiles, weighted by their respective mass. This representation effectively captures the expected frequency with which each strategy is employed by the players in the long run.

The ranking process is driven by the conditional switch rate $\rho$, defined by a selection function. Specifically, the conditional switch rate \(\rho_{\sigma, \tau}^i(s_{-i})\) describes the probability of player $i$ in a population switching from strategy \(\sigma\) switches to strategy \(\tau\) , given the strategy choices \(s_{-i}\) of other players, is given by

\begin{equation}\label{eq:probability}
\rho_{\sigma, \tau}^i (s_{-i})=\dfrac{1 - e^{-\alpha(u_i(\tau, s_{-i}) - u_i(\sigma, s_{-i}))}}{1 - e^{-N\alpha(u_i(\tau, s_{-i}) - u_i(\sigma, s_{-i}))}},
\end{equation}
if $u_i(\tau, s_{-i}) \neq u_i(\sigma, s_{-i})$, and $\rho_{\sigma, \tau}^i (s_{-i})=\frac{1}{N}$, otherwise, where $N$ is the number of players in the population, and $u_i(\tau, s_{-i})$ and $ u_i(\sigma, s_{-i})$ represents player $i$'s payoffs using strategies $\tau$ and $\sigma$, respectively. The parameter $\alpha$, known as the \emph{ranking intensity}, controls the selection strength. To ensure that even small differences in payoffs contribute to a player's strategy switch, and to guarantee a unique stationary distribution, $\alpha$ must be sufficiently large. \par
In the first game, where players share uniform latency and orderflow access distribution, we begin with a small $\alpha$ value and increase it exponentially (i.e., \emph{sweep}), following the approach suggested in \cite{alpharank}. However, as the size of the game grows, this sweeping becomes computationally challenging. Thus, we use a method from OpenSpiel \cite{LanctotEtAl2019OpenSpiel} to estimate a lower-bound value of $\alpha$.\footnote{\url{https://github.com/google-deepmind/open_spiel/pull/403}.}

\section{Results}
\label{sec:results}
In this section, we present the results of the three games. We examine the equilibria of these games, focusing on the strategic incentives of builders under different conditions. We analyze how advantages in latency and orderflow access influence builders' incentives, and assess the overall state of the MEV-Boost auction. We discuss how our findings compare to those in \cite{icaifpaper} in \Cref{sec:discussion}.

\subsection{Symmetric game}
\begin{figure}[t]
\centering
\includegraphics[width=0.85\linewidth]{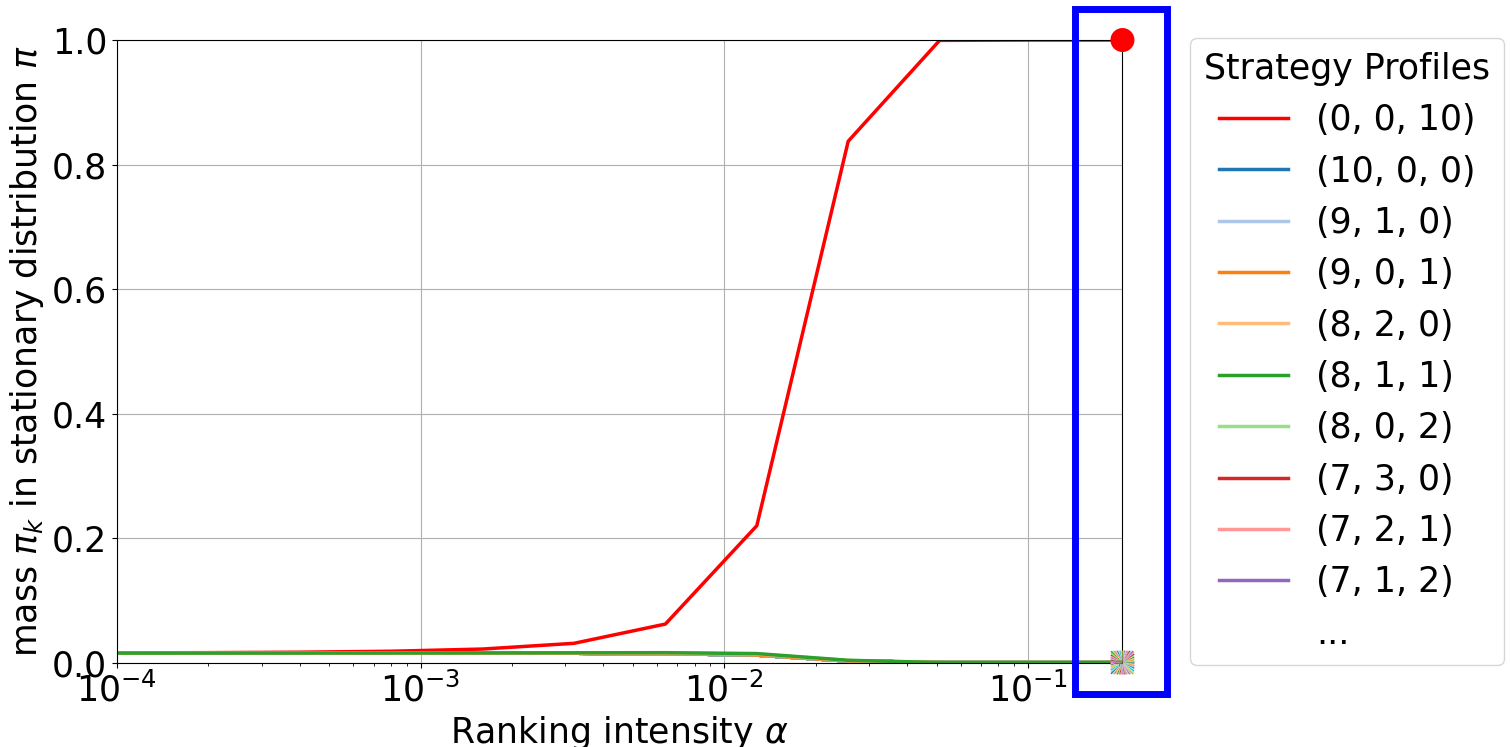}
\caption{Mass of all strategy profiles in the stationary distribution when sweeping $\alpha$. The legend shows the final ranking of strategy profiles as computed by $\alpha-$Rank.}
\label{fig:alpha_sweep}
\end{figure}

In the game where all 10 players have uniform latency and the same prior distribution of private orderflow access probability, the strategy profile in which all 10 players adopt the aggressive bidding strategy demonstrates clear dominance, with a stationary probability of 1, while all other strategy profiles have a stationary probability of 0. Figure~\ref{fig:alpha_sweep} shows the equilibrium as computed by $\alpha$-Rank. This result indicates that the game converges to a \emph{pure Nash equilibrium}.

This equilibrium can be classified as a \emph{symmetric Bayesian Nash equilibrium}, where all builders adopt the same strategy and no one has the incentive to deviate under these conditions. Specifically, all 10 builders, with uniform latency and comparable private orderflow access, are incentivized to compete by aggressively bidding a large portion of their private signal in the auction. This results in an equal win rate of 10\% for each builder in expectation. While it is theoretically possible for the builders to secure the same 10\% win rate with higher payoffs by collectively adopting the conservative strategy—bidding only a small portion of their private signal, effectively engaging in a form of \emph{collusion}—this strategy is not incentive-compatible and the strategy profile is not evolutionary stable. If any builder deviates by bidding more aggressively, they would significantly increase their chances of winning the auction, prompting other builders to follow suit and bid more aggressively.

These findings suggest that, under idealized conditions—where builders experience equal latency and have comparable private orderflow access—the current MEV-Boost auction mechanism effectively incentivizes builders to compete by bidding a large portion of their private signal. This results in an efficient auction mechanism, with the majority of the MEV being captured by the proposer through the auction. Furthermore, it fosters a decentralized market where builders share the same market share (i.e., win rate) in expectation.

\subsection{Latency impact}
In the first role-symmetric game, all players maintain the same prior distribution of private orderflow access, but experience two different types of latencies (low and high). The key factor affecting builders' strategy choices is the latency gap between low- and high-latency groups. To analyze this, we consider scenarios where the latency difference between the two groups ranges from 0ms (previous symmetric scenario) to 200ms, increasing in 10ms increments. The latency of low-latency players remains fixed at 10ms, while the latency of high-latency players starts at 10ms and increases by 10ms in each subsequent scenario. Each scenario is treated as an independent game. We analyze the equilibria of these games to assess the impact of latency differences.\par

\begin{figure}[t]
\centering
\includegraphics[width=0.49\linewidth]{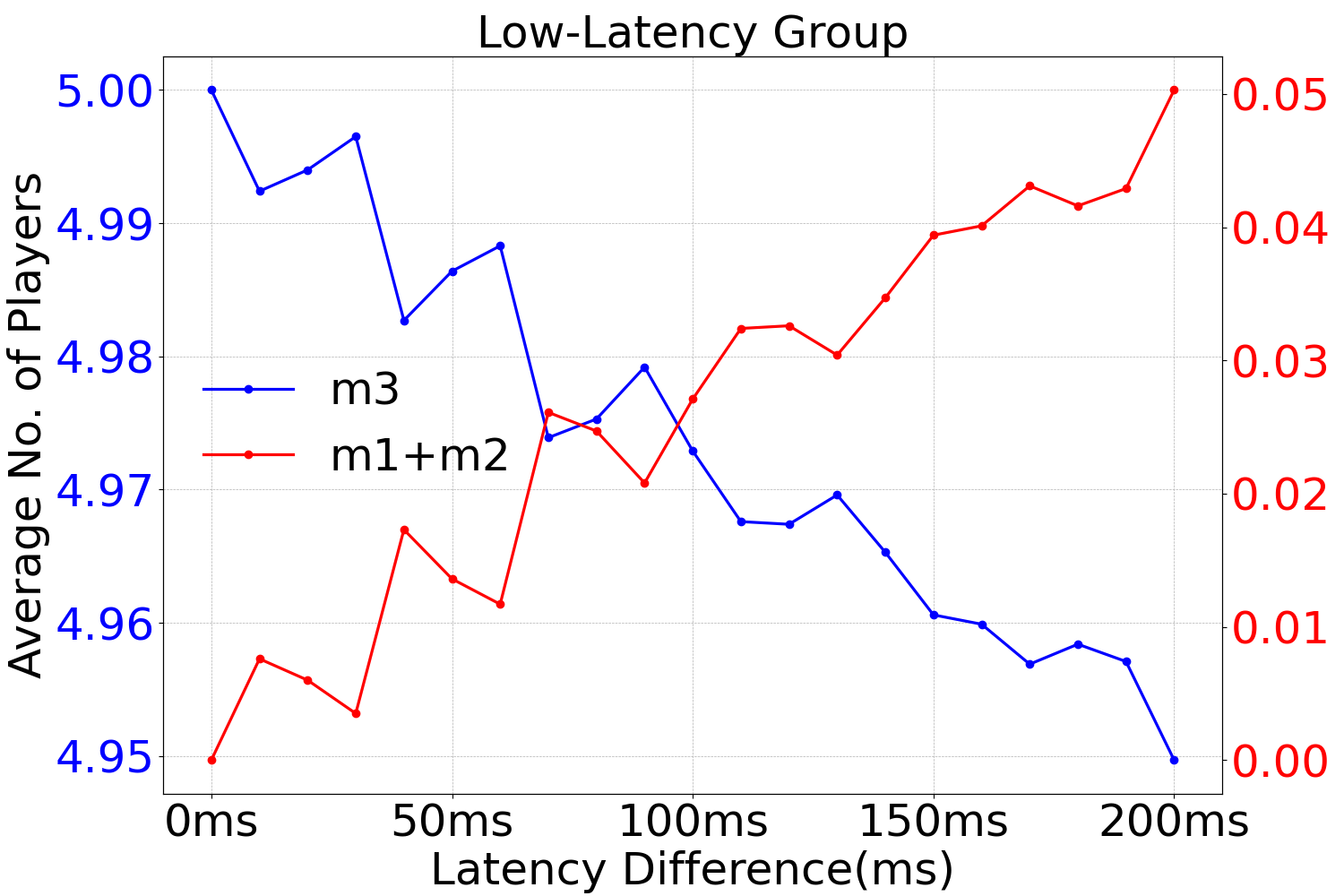}
\includegraphics[width=0.49\linewidth]{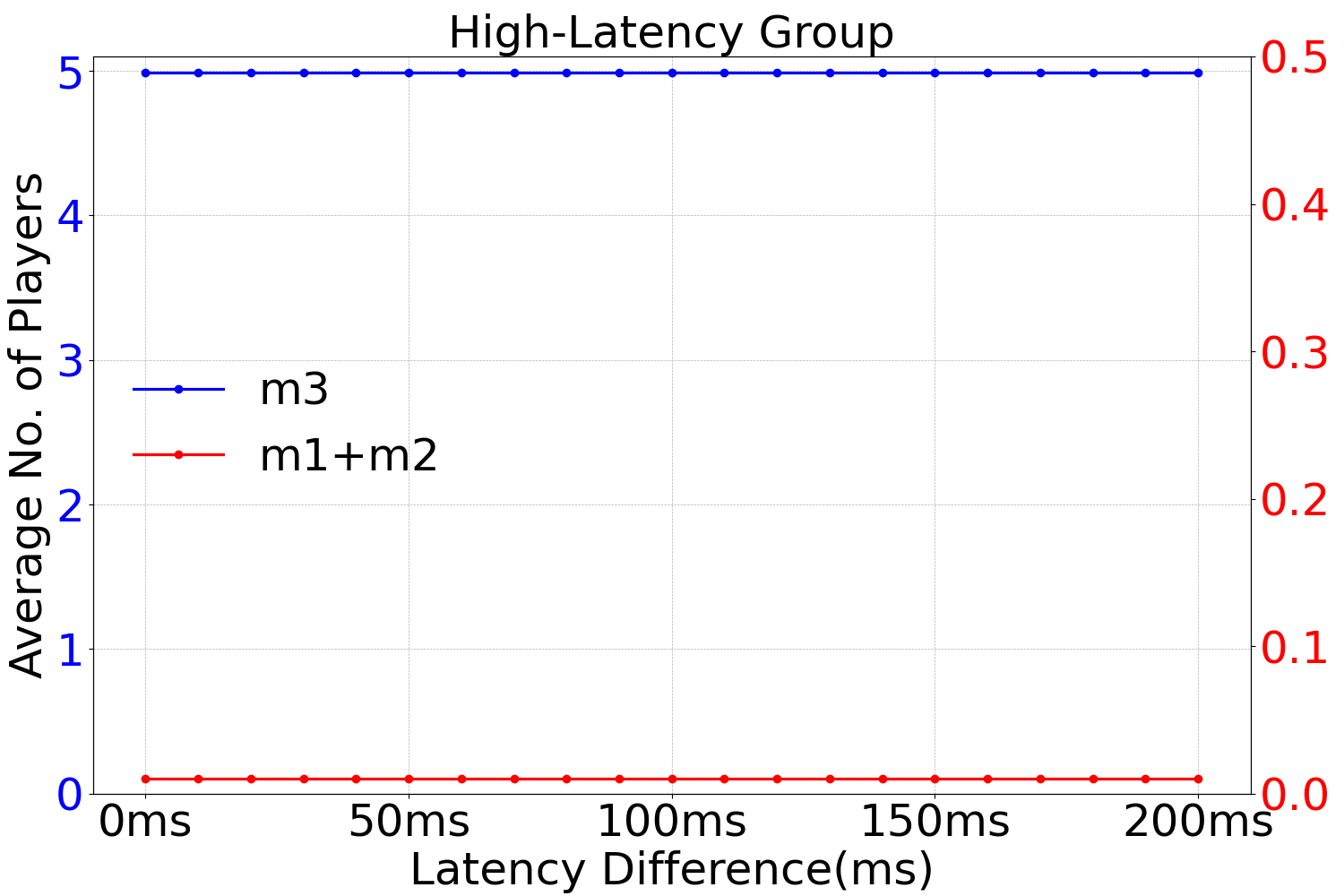}
\caption{Average usage of aggressive and the other two (conservative+moderate) strategies by low-latency players (left) and high-latency players (right) across all profiles under varying latency differences as computed by $\alpha$-Rank.}
\label{fig:eq_latency}
\end{figure}

It turns out that the latency gap between the two groups does not significantly affect players' incentives compared to the previous symmetric scenario. As shown in Figure~\ref{fig:eq_latency}, both the low- and high-latency groups remain strongly incentivized to adopt the aggressive bidding strategy in equilibrium under various latency differences. High-latency players must still bid aggressively to counter their latency disadvantage. On the other hand, low-latency players may sometimes win with moderate or conservative bids leading to a slight increase in those strategies as latency differences grow. However, this only occurs in certain edge cases when high-latency players miss valuable transactions near the auction's end, and the overall equilibrium remains largely stable with aggressive bidding prevailing across both groups. 

While the incentives of both groups remain largely unchanged, the latency gap does shift market dynamics, favoring low-latency players. To quantify this, we calculate the \emph{overall win rates} of both latency groups at equilibrium, derived as the average of the win rates across all strategy profiles, indexed by $k$, and weighted by their stationary probabilities, $\pi_k$. Formally, for a profile $m^k$, let the win rate of the low-latency group be denoted by $w^{l}_{k}$, and the win rate of the high-latency group $w^{h}_{k}$. The overall win rates for the low-latency group and high-latency group at equilibrium are:
\begin{align}
w^{l} = \sum\nolimits_{k} \pi_k \cdot w^{l}_{k}, \text{ and }
w^{h} = \sum\nolimits_{k} \pi_k \cdot w^{h}_{k}.
\end{align}

\begin{figure}[t]
\centering
\includegraphics[width=0.49\linewidth]{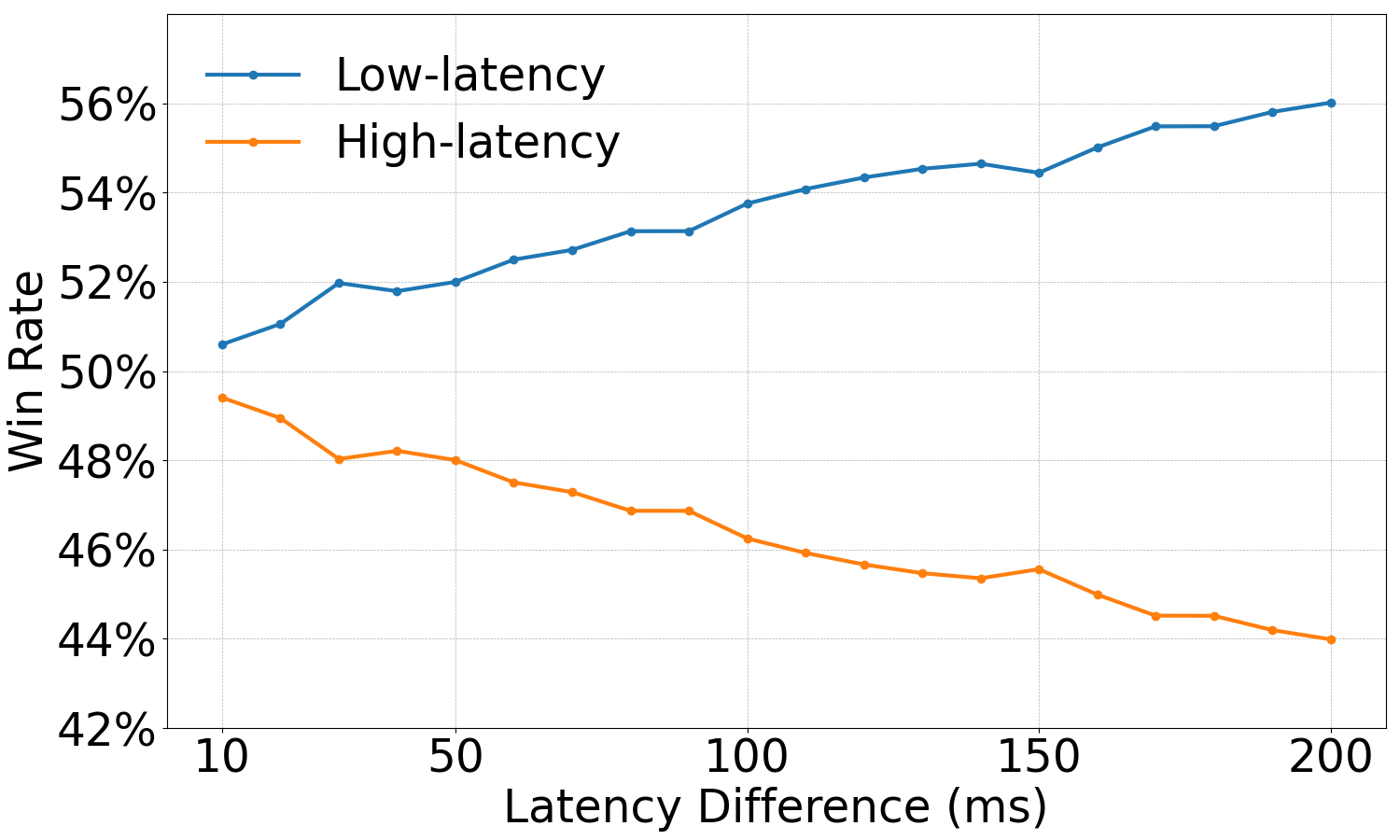}
\includegraphics[width=0.49\linewidth]{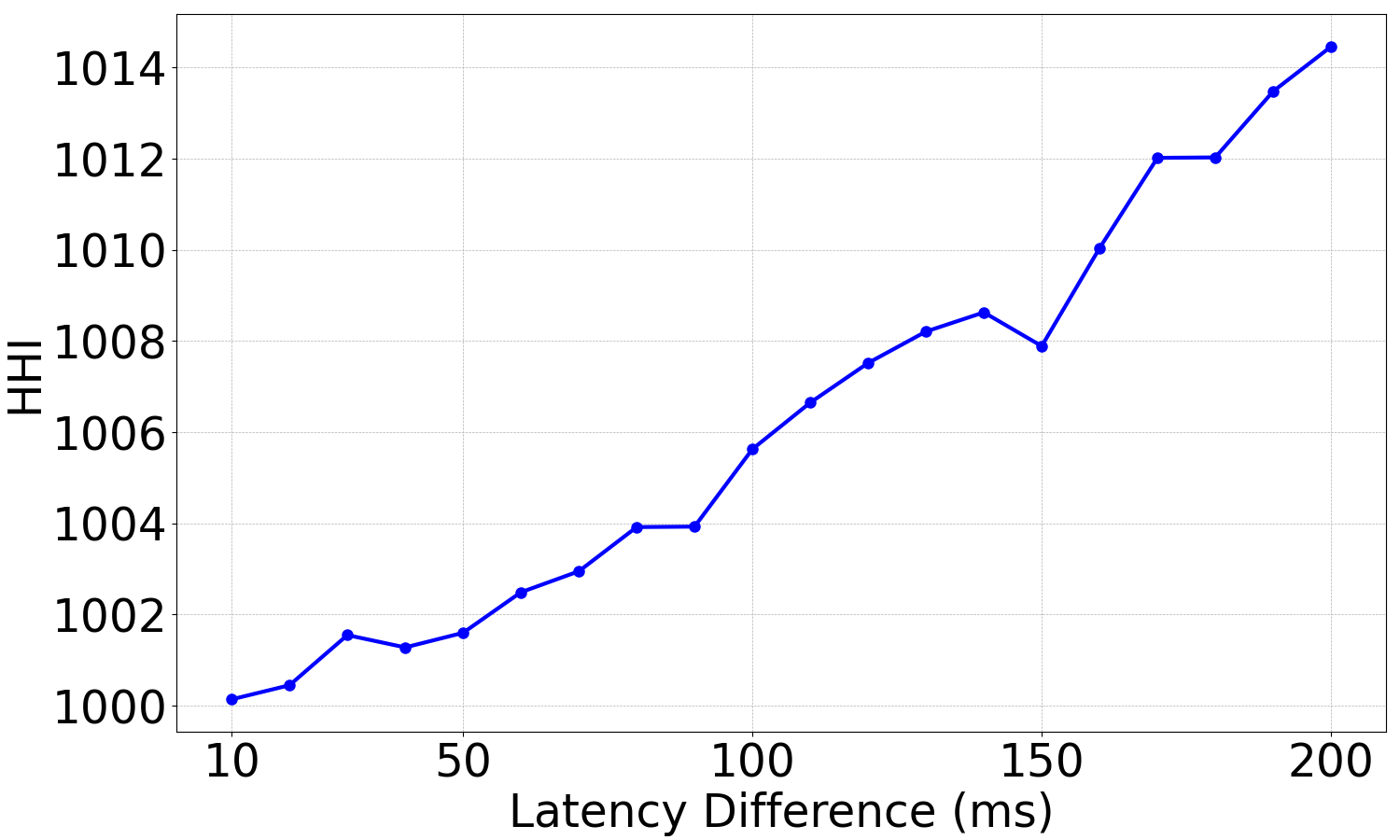}
\caption{Overall win rates (left) of low- and high-latency players and HHI (right) under varying latency differences.}
\label{fig:latency_winrate}
\end{figure}

We further compute the Herfindahl-Hirschman Index (HHI) to measure the level of market centralization at equilibrium \cite{presto}. The win rate of each individual player is assumed to be evenly distributed within their respective groups. For each group, the individual win rates are: 
\begin{equation}
w_{i}^{l} = \dfrac{w^{l}}{5}, \quad w_{i}^{h} = \dfrac{w^{h}}{5}.
\end{equation}
Finally, the HHI is computed as the sum of the squared individual win rates for all 10 players:
\begin{equation}\label{eq:HHI}
HHI = \sum_{i=1}^{5}\left(w^{l}_i\right)^2 + \sum_{i=1}^{5}\left(w^{h}_i\right)^2 =\frac{1}{5} \left(\left(w^{l}\right)^2 +\left(w^{h}\right)^2\right).
\end{equation}
The left plot in Figure~\ref{fig:latency_winrate} shows that low-latency players secure higher win rates and capture larger market shares as their latency advantage grows, with differences in the win rates reaching approximately 12\%. However, our findings suggest that while improved latency can result in better performance, latency differences are \emph{not} the primary driver of centralization. The right plot presents the HHI under varying latency differences between low- and high-latency players, showing that despite a trend toward centralization, the market remains relatively competitive, with an HHI below 1500. These results indicate that the latency gap alone—typically no greater than 200ms between low- and high-latency builders—is insufficient to significantly centralize the market. Nonetheless, builders are incentivized to vertically integrate with orderflow providers and relays for latency optimization, targeting faster access to orderflow and bid updates, thereby contributing to a gradual shift toward centralization.


\begin{figure}[t]
\centering
\includegraphics[width=0.75\linewidth]{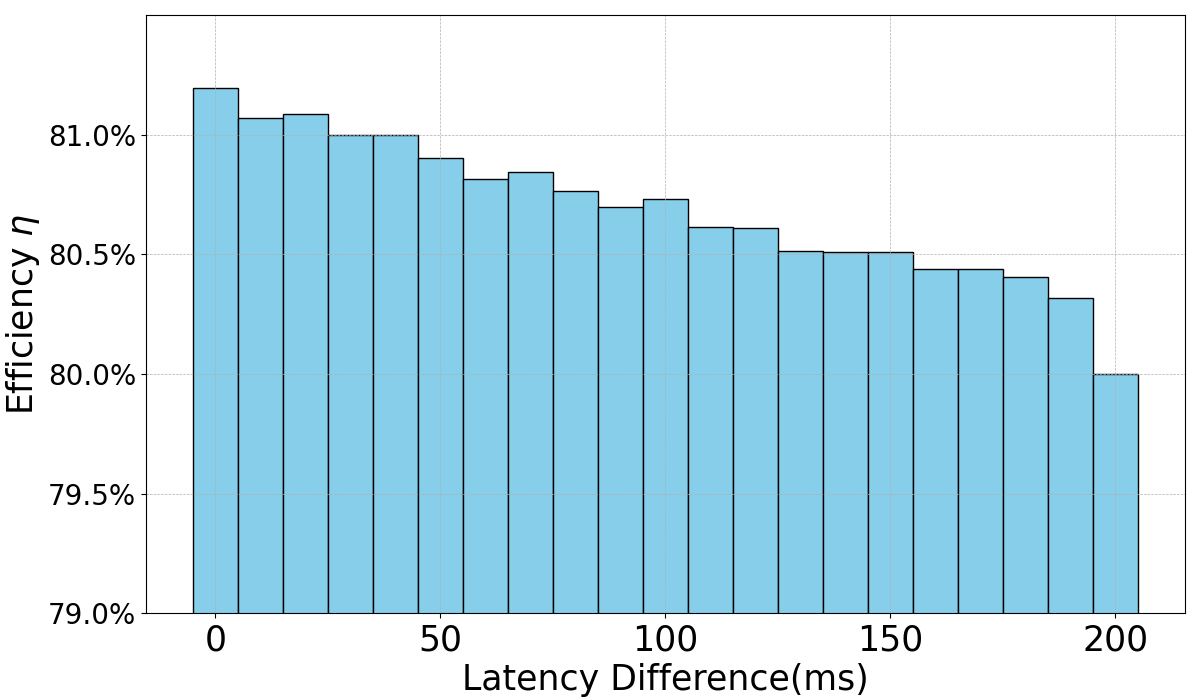}
\caption{The overall auction efficiencies under varying latency differences.}
\label{fig:latency_eff}
\end{figure}

The latency disparity between builders can have a more significant impact in decreasing the proposer revenue, i.e., winning bid value, and reduce auction efficiency. This typically occurs when high-latency players win auctions but miss signal updates near the end due to their latency—meaning they could have incorporated more signal if their latency were lower. Additionally, as described above, low-latency players may win by bidding a smaller portion of their private signal when they have a significant latency advantage, resulting in less MEV captured through the auction. To evaluate the MEV-Boost auction mechanism's capability to capture MEV, we introduce the metric of \emph{efficiency}, defined as the ratio of the winning bid value to the total signal. Formally, we denote with $\eta_{k,v}$ the auction efficiency for strategy profile $m^k$ in simulation round $v \in \{1,2,\dots,1000\}$. The average efficiency for the strategy profile $m^k$ over its 1,000 simulations is, thus, given by: 
\begin{equation}
    \overline{\eta}_{k} := \dfrac{1}{1000} \sum\nolimits_{v=1}^{1000} \eta_{k,v}.
\end{equation} 
The \emph{overall efficiency}, denoted by $\eta$, which captures the auction efficiency at equilibrium, is computed by combining the average efficiency $\overline{\eta_{k}}$ for each profile $m^k$ and weighing it by its stationary probability $\pi_k$: 
\begin{equation}
    \eta = \sum\nolimits_{k}\pi_k \cdot \overline{\eta}_{k}.
\end{equation}
As is shown in Figure~\ref{fig:latency_eff}, the overall auction efficiency decreases as the latency difference between the low- and high-latency players increases. This is because low-latency players are more likely to win the auction by incorporating a smaller portion of their private signal into their bid as the latency disadvantage of high-latency players grows.

Although further increasing the latency difference between the low- and high-latency players can affect the players' incentive, particularly for low-latency players, and eventually shift the equilibrium, empirical studies \cite{whowinsandwhy} have shown that, among the top ten builders, the average latency difference between the five faster builders and five slower builders is typically around 200ms. This indicates that the setup of our games provides practical insights into the real market.

\subsection{Orderflow access impact}
Next, we analyze builders' incentives when their orderflow access probabilities differ, i.e., they are not randomly drawn from the same prior uniform distribution. As described in Section~\ref{sec:games}, we mirror the approach used in the latency impact analysis and examine scenarios involving 5 players with low orderflow access probability and 5 players with high orderflow access probability. We refer to them as low-orderflow and high-orderflow players, respectively. Specifically, the orderflow access probability for low-orderflow players is fixed at 30\% (i.e., the minimal threshold), while the probability for high-orderflow players starts at 40\% and increases by 10\% in each subsequent scenario, reaching up to 100\%. Again, each scenario of probability difference is treated as an independent game and we analyze the equilibria of these games to understand the impact of orderflow access on high- and low-orderflow players' bidding incentives.\par

\begin{figure}[t]
\centering
\includegraphics[width=0.49\linewidth]{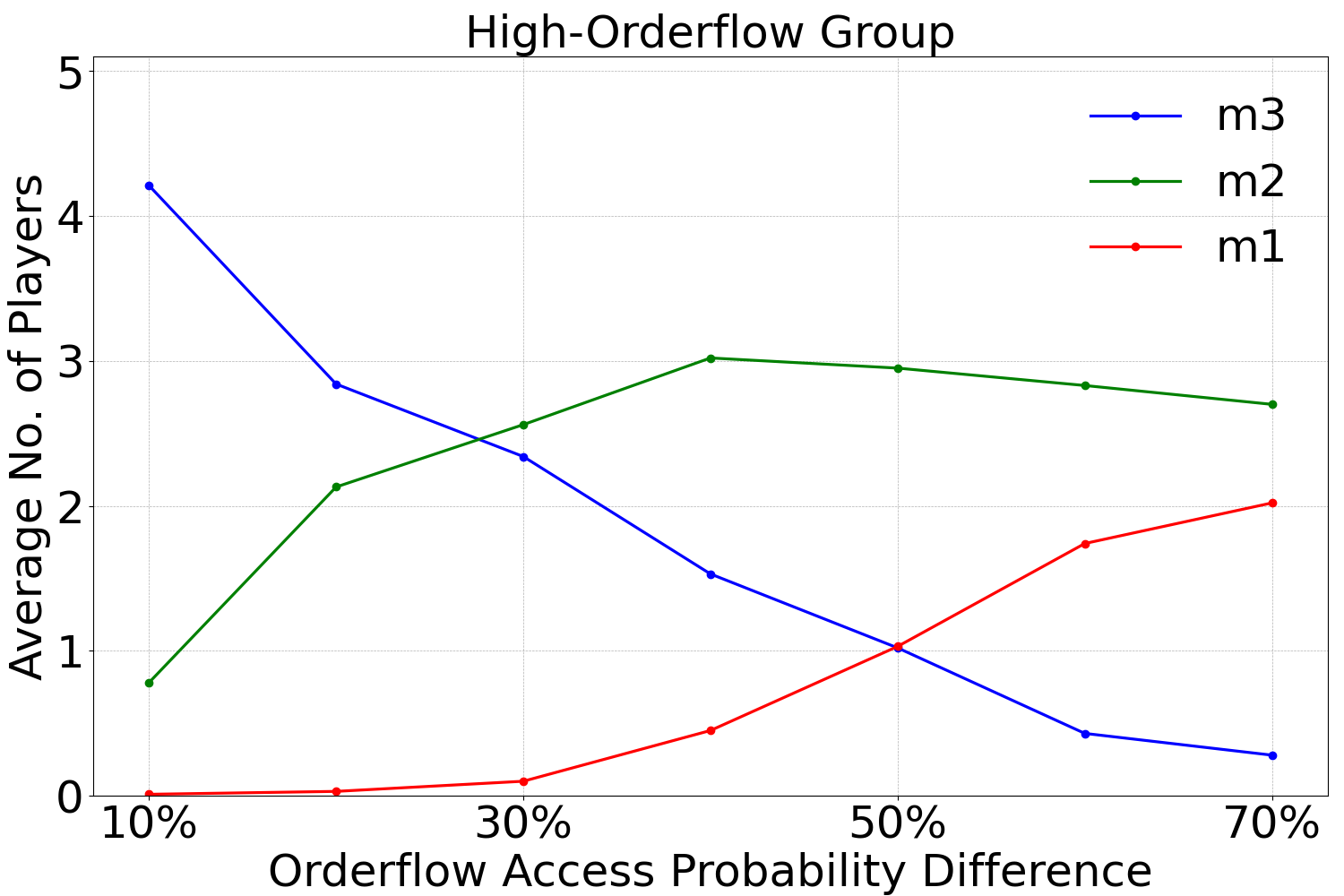}
\includegraphics[width=0.49\linewidth]{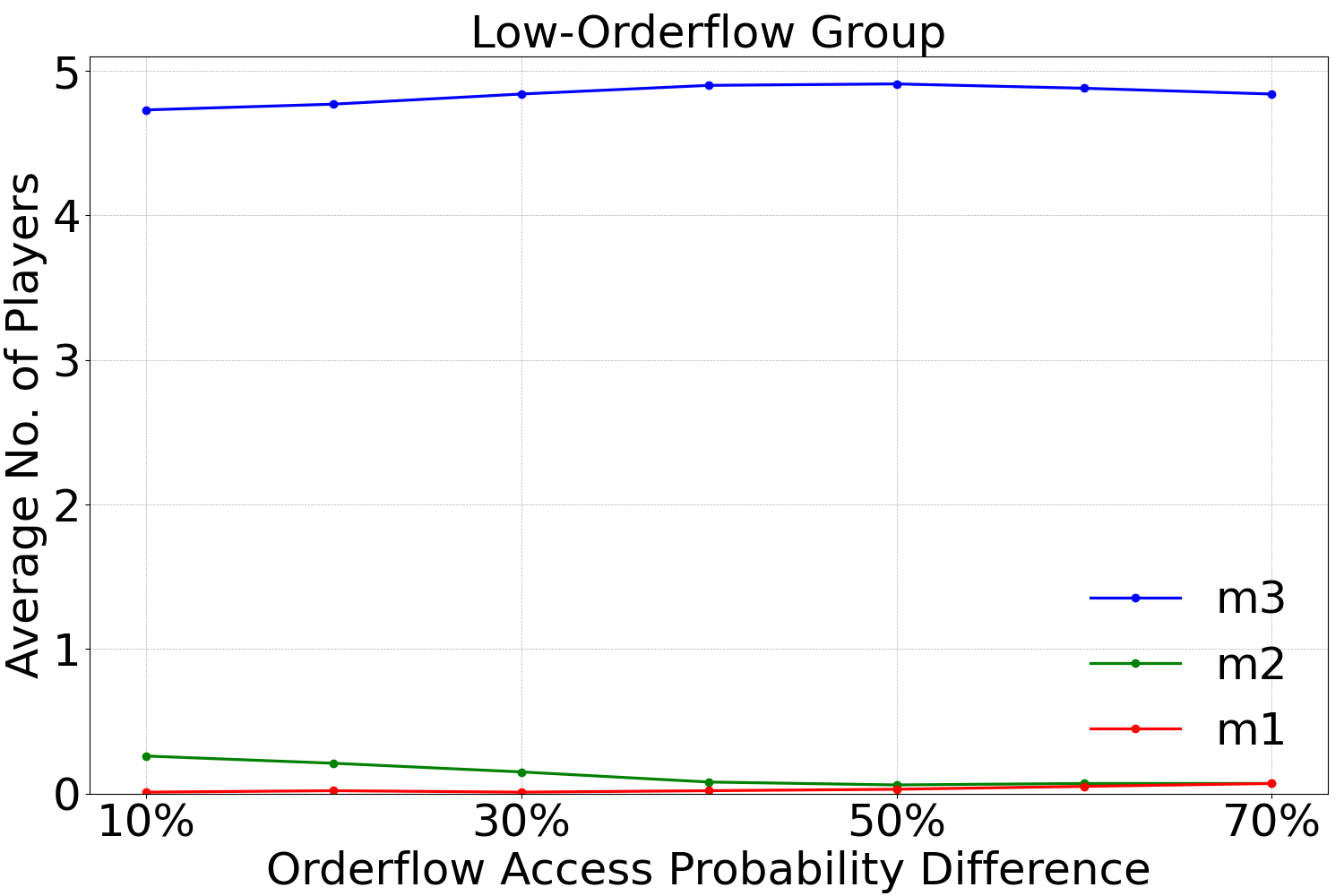}
\caption{Average usage of each meta strategy by high-orderflow players (left) and low-orderflow players (right) across all profiles under varying orderflow access probability differences as computed by $\alpha$-Rank.}
\label{fig:eq_orderflow}
\end{figure}

Figure~\ref{fig:eq_orderflow} presents the equilibria of the above games. As high-orderflow players' access probability rises, they can win auctions by bidding a smaller portion of their private signal. This reduces their incentive to bid aggressively, prompting a shift towards moderate and even conservative strategies to maximize profits. When the gap in orderflow access widens significantly, high-orderflow players can secure wins with minimal bids.\par

Conversely, low-orderflow players remain strongly incentivized to bid aggressively. When the orderflow access probability gap is small, they still have occasional chances to win with moderate strategy. However, as the probability difference grows, with limited access to orderflow, they must use most of their private signal to stand a chance of winning. Interestingly, we observe a slight increase in the use of moderate and conservative strategies by low-orderflow players when the probability gap becomes significant. This may be due to the increasing adoption of conservative strategies by high-orderflow players, which creates random opportunities for low-orderflow players to win with less aggressive strategies.

\begin{figure}[t]
\centering
\includegraphics[width=0.49\linewidth]{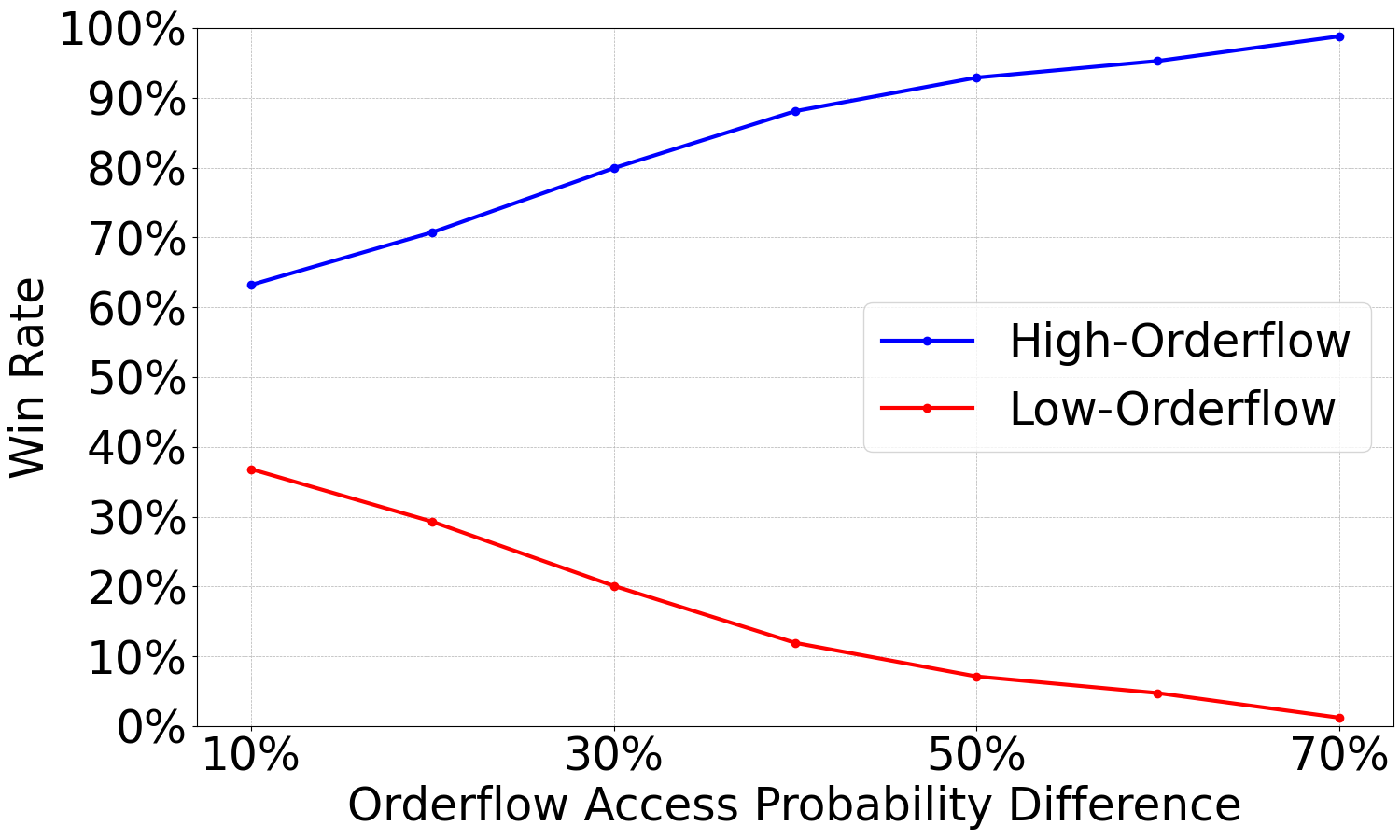}
\includegraphics[width=0.49\linewidth]{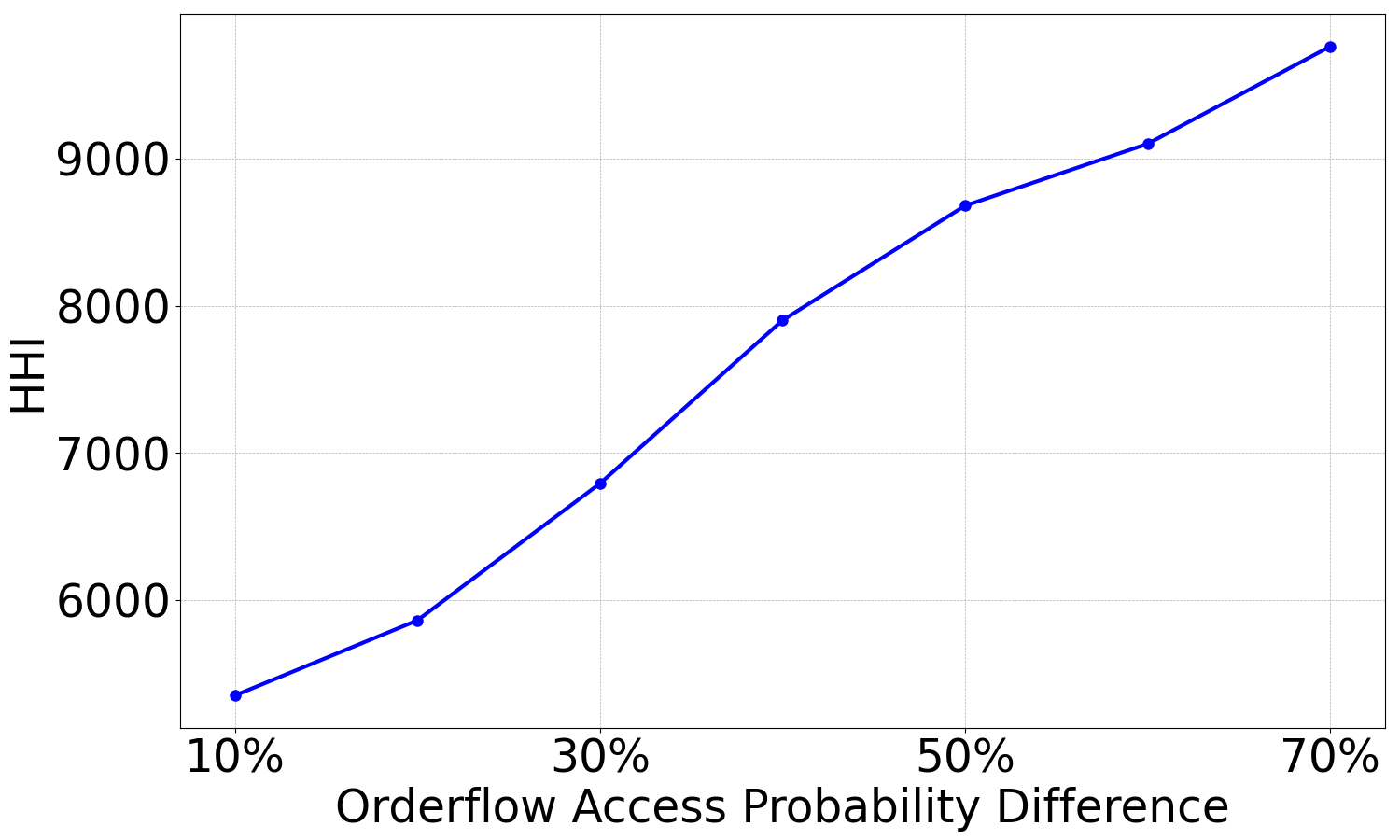}
\caption{Overall win rates (left) of low- and high-orderflow players and HHI (right) under varying orderflow access probability differences.}
\label{fig:orderflow_winrate}
\end{figure}

\begin{figure}[t]
\centering
\includegraphics[width=0.8\linewidth]{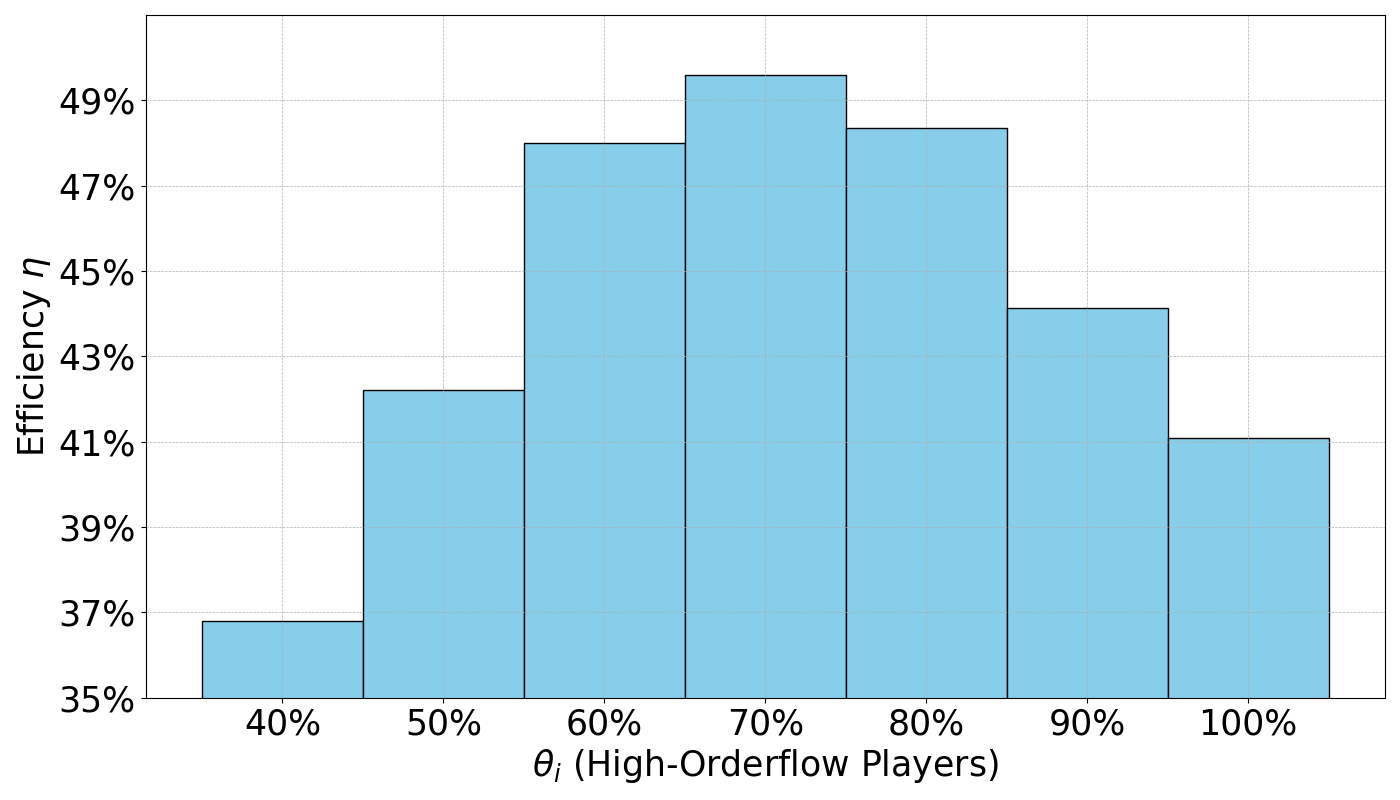}
\caption{The overall auction efficiencies under varying orderflow access probabilities of high-orderflow builders.}
\label{fig:eff_orderflow}
\end{figure}

As in the latency impact analysis, we calculated the overall win rates for high- and low-orderflow groups and the HHI under varying orderflow access probability differences. The results are shown in Figure~\ref{fig:orderflow_winrate} and reveal the strong oligopolistic dynamic driven by disparities in orderflow access in the Ethereum block building market. Builders with superior access, especially those who vertically integrate with orderflow providers, gain a competitive edge, dominating the market by bidding conservatively yet still winning blocks. This allows them to secure blocks with lower bids while maintaining high profit margins. Such oligopolistic behavior results in reduced proposer revenue and lower auction efficiency. Figure~\ref{fig:eff_orderflow} highlights this effect. Initially, as the orderflow access of high-orderflow players increases, there is an improvement in auction efficiency, since builders incorporate more MEV into their bids. However, when high-orderflow players dominate the orderflow landscape, their incentive shifts toward moderate and conservative bidding. These bidding strategies decrease the proportion of MEV captured by proposers, ultimately lowering the overall auction efficiency.\par
As these high-orderflow builders win more auctions and expand their market share, they benefit from an \emph{economy of scale}. Orderflow providers, aiming to maximize the probability of their transactions landing on-chain, prioritize sending more orderflow to dominant builders. This creates a \emph{reinforcing feedback loop}: the more blocks a builder wins, the more orderflow they receive, which further strengthens their ability to dominate future auctions. When combined with latency advantages, which are commonly enjoyed by high-orderflow builders (e.g., \texttt{beaverbuild} and \texttt{Titan}), this further amplifies their dominance, leading to even lower auction efficiency. This growing advantage allows high-orderflow builders to entrench their dominance while enjoying higher profit margins. 

To counter this trend, smaller builders, with limited access to private orderflow, are often forced to bid aggressively, making minimal profits or even subsidizing bids to maintain their presence in the market \cite{whowinsandwhy}, which is financially unsustainable in the long term. This observed dynamic raises concerns about the long-term health of the market, as it discourages new entrants and entrenches the position of established builders, contributing to further centralization of the Ethereum block building market \cite{yang2024decentralization}.

\section{Discussion}
\label{sec:discussion}
Our results show that under idealized conditions --where all builders share equal latency and orderflow access-- the block building market remains competitive and decentralized, with an efficient MEV-Boost auction. 
However, in practice, exclusive orderflow deals arise largely because orderflow providers have no trustless, private way to distribute flow; they form exclusive agreements with builders to protect their MEV and secure priority block inclusion. Consequently, large builders receive preferential orderflow, while smaller builders face limited access and struggle to win. Vertically integrating with MEV searchers and relays further grants certain builders more frequent and valuable orderflow, enabling them to bid conservatively yet maintain high margins. These advantages drive centralization and reduce auction efficiency.

Recent proposals like \texttt{MEV-Share} \cite{mevshare}, \texttt{MEVBlocker} \cite{mevblocker}, and \texttt{BuilderNet} \cite{buildernet} aim to provide a fairer and private orderflow distribution. However, these systems are either relying on trusted entities or remain in early development, leaving their long-term impact on market structure unclear.

\paragraph{Robustness of results.}Unlike \cite{icaifpaper}, where their findings are inherently constrained by their restricted strategy space and not observed in real-world dynamics \cite{whowinsandwhy, yang2024decentralization}, the main advantage of the meta-game framework is that it allows us to explore the strategy space in a more nuanced way while keeping the analysis tractable. In our case, this is achieved through the analysis of three \emph{meta-strategies} that partition the entire strategy space. Despite the complexity of real-world MEV-Boost auctions, this model is rich enough to recover emerging behaviors and provides meaningful insights into builder incentives, auction efficiency, and market (de)centralization. Importantly, our current set of results is robust to the selected partition of the strategy space; our tests (not presented here) suggest that finer partitions do not qualitatively change the results while significantly increasing the analytical complexity.

\section{Conclusion}
In this paper, we examined builders' strategic bidding incentives in Ethereum block building auctions through empirical game theoretic analysis within the meta-game framework. Our results show that under idealized market conditions 
builders are incentivized to bid aggressively, resulting into a competitive and decentralized market with efficient auctions. However, when builders have unequal resources, as is the case in practice, the market tends to centralize. Well-resourced builders, especially those with superior access to private orderflow, are favored to benefit from economy of scale. This creates an oligopolistic dynamic, leading to reduced proposer revenue and lower auction efficiency. Our results highlight the importance of fair MEV distribution among builders and underscore the necessity of enhancing decentralization in the Ethereum block building market. \par
From a multi-agent system perspective, our results show how the meta game framework, allowing to account for an infinite number of strategies, qualitatively changes the findings of previous literature \cite{icaifpaper} in the domain: In MEV boost auctions, builders do not collude, but instead compete to create an oligopoly. 

\begin{ack}
We thank Julian Ma of Ethereum Foundation and one AFT reviewer for their insightful comments on the early version of this work.
\end{ack}


\bibliography{mybibfile}

\end{document}